\documentclass[]{raa}
\usepackage{cases}            
\usepackage{graphicx,times,subfig}
\usepackage{natbib}
\usepackage{longtable,textcomp}

\providecommand{\bm}[1]{\mbox{\boldmath$#1$\unboldmath}}

\begin{document}

   \title{A statistical classification of the unassociated gamma-ray sources in the second Fermi Large Area Telescope Catalog
}

   \volnopage{Vol.0 (201x) No.0, 000--000}      
   \setcounter{page}{1}           

   \author{Zhu Mao
   \and Yun-Wei Yu\mailto{}
%
      }

   \institute{Institute of Astrophysics, Central China Normal
University, Wuhan 430079, China \email{yuyw@phy.ccnu.edu.cn}
          }

\date{Received~~2013 month day; accepted~~2013~~month day}

\abstract{With assistance of the identified/associated sources in
the second Fermi Large Area Telescope (LAT) catalog, we analyze and
resolve the spatial distribution and the distributions of the
gamma-ray spectral and variability indices of the remaining 575
unassociated Fermi LAT sources. Consequently, it is suggested that
the unassociated sources could statistically consist of Galactic
supernova remnants/pulsar wind nebulae, BL Lacertae objects, flat
spectrum radio quasars, and other types of active galaxies with
fractions of 25\%, 29\%, 41\%, and 5\%, respectively.
\keywords{gamma rays: general - catalogs - methods: statistical }}

   \authorrunning{Z. Mao \& Y. W. Yu}            
   \titlerunning{A statistical classification of the unassociated Fermi LAT sources}  

   \maketitle

%
%
\section{INTRODUCTION}

A routine whole-sky survey in the 100 MeV to 100 GeV band with the
Large Area Telescope (LAT; Atwood et al. 2009) on board the {\it
Fermi Gamma-Ray Space Telescope} has been carried out since the
science phase of the mission began in August 2008. By analyzing the
observational results of the first 24 months, the Fermi LAT
Collaboration published the second Fermi LAT Source Catalog
containing 1873 gamma-ray sources (Nolan et al. 2010). Great efforts
have been made to identify the LAT sources by periodic emission or
variability correlation with other wavelengths, and also to provide
associations for much more LAT sources with previous gamma-ray
catalogs and with likely counterpart sources from known or suspected
source classes (based on Bayesian probabilities in a LAT error box).
As a result, 1298 LAT sources are identified as or associated to
pulsars, supernova remnants (SNRs), pulsar wind nebulae (PWNe),
blazars, active galaxies, etc (see Table 1). However, there are
still nearly one third of the LAT sources unclassified (i.e., 575
out from 1873). Such a situation is similar to that of the third
{\it Energetic Gamma Ray Experiment Telescope} (EGRET) catalog
(Hartman et al. 1999).
\begin{center}
\begin{table}
\caption{The numbers of the identified/associated sources in the
second Fermi LAT catalog}
\begin{center}
\begin{tabular}{lcc}\hline\hline\noalign{\smallskip}
Types& Nos. \\
\hline\noalign{\smallskip}
 Pulsar     & 108 \\           
 Pulsar wind nebula                                   & 3  \\           
Supernova remnant                                    & 10 \\          
 Supernova remnant/Pulsar wind nebula                & 58 \\          
 Globular Cluster                                     & 11  \\           
 High-mass binary                                     & 4   \\          
Nova                                                 & 1   \\          
 \hline
 BL Lacerate object                                & 436 \\          
 Flat spectrum radio quasar                                  & 370  \\         
 Active galaxy of uncertain type                      & 257  \\          
 Non-blazar active galaxy                             & 11  \\          
 Radio galaxy                                         & 12  \\          
 Seyfert galaxy                                       & 6  \\           
 Normal galaxy                                        & 6   \\          
 Starburst galaxy                                     & 4   \\          
 \hline\hline
\end{tabular}

\end{center}
\end{table}
\end{center}

The difficulty in the identification/association of these
unclassified sources is because of that the location accuracy of the
LAT sources is typically insufficiently precise. In other words, in
a typical LAT error box, there are too many stars, galaxies, X-ray
sources, infrared sources, and radio sources. Therefore, besides the
source positions, more information is required to determine the
nature of the unassociated LAT sources, including the spectral
information, the time variability, and the availability of a
plausible physical process at the source to produce
sufficiently-high energy gamma-rays. Stephen et al. (2010) made such
an attempt with the {\it ROSAT} catalog which is useful in finding a
positionally corrected, highly unusual object that might be expected
to produce gamma-rays. As a result, 30 unassociated LAT sources are
tentatively suggested to be associated with a {\it ROSAT} countpart.
More directly, Elizabeth et al. (2012) proposed that, a comparison
of the spectral and variability indices between the associated and
unassociated LAT sources can provide insight into the likely classes
of the unassociated sources. Following such a consideration, in this
paper, we would give a statistics on the gamma-ray spectral and
variability indices of the unassociated Fermi LAT sources. Based on
these statistics, we try to find out that how many components the
unassociated sources consist of and what these components could be.
\begin{figure}
\centering
\resizebox{\hsize}{!}{\includegraphics{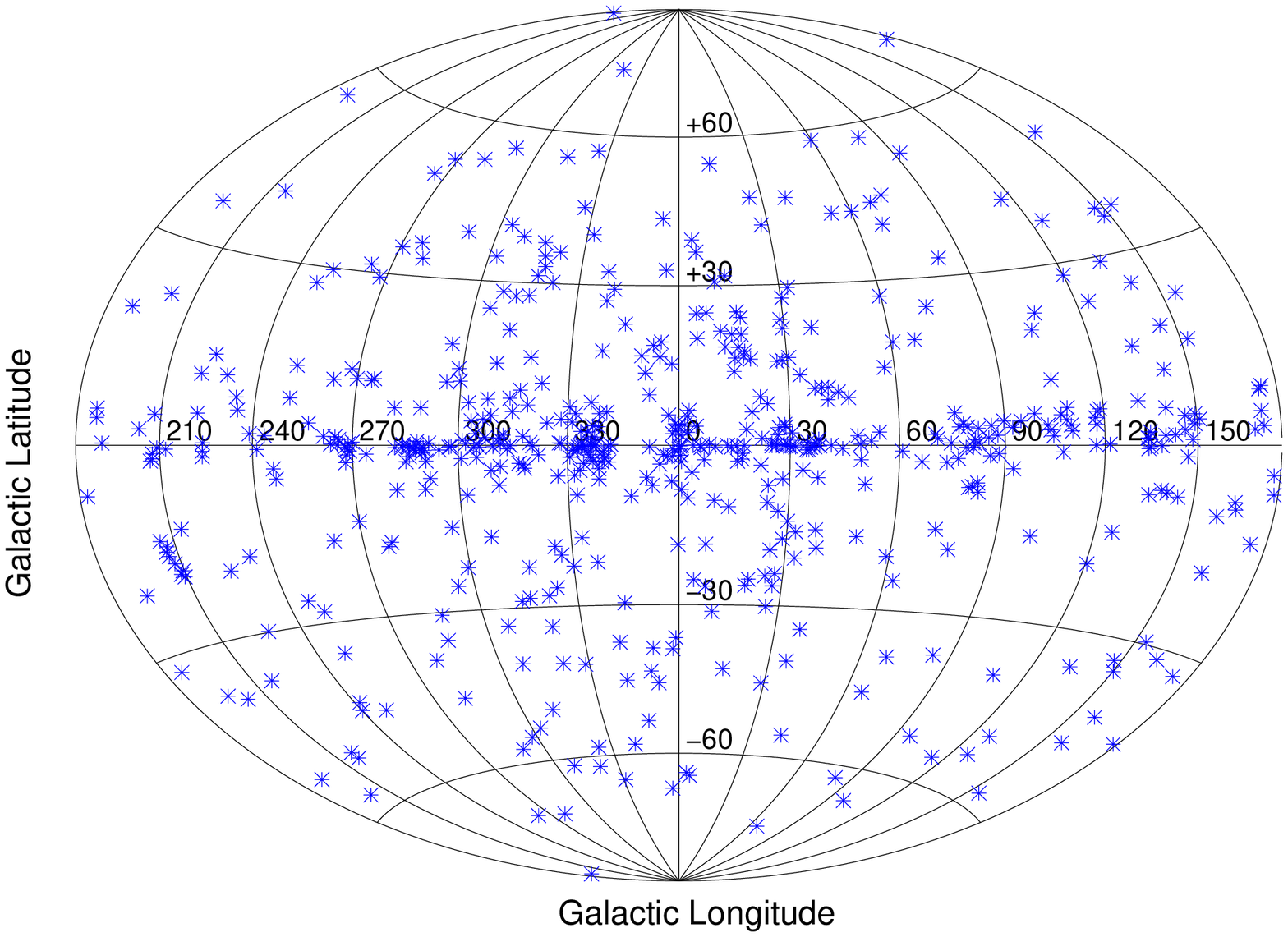}\includegraphics{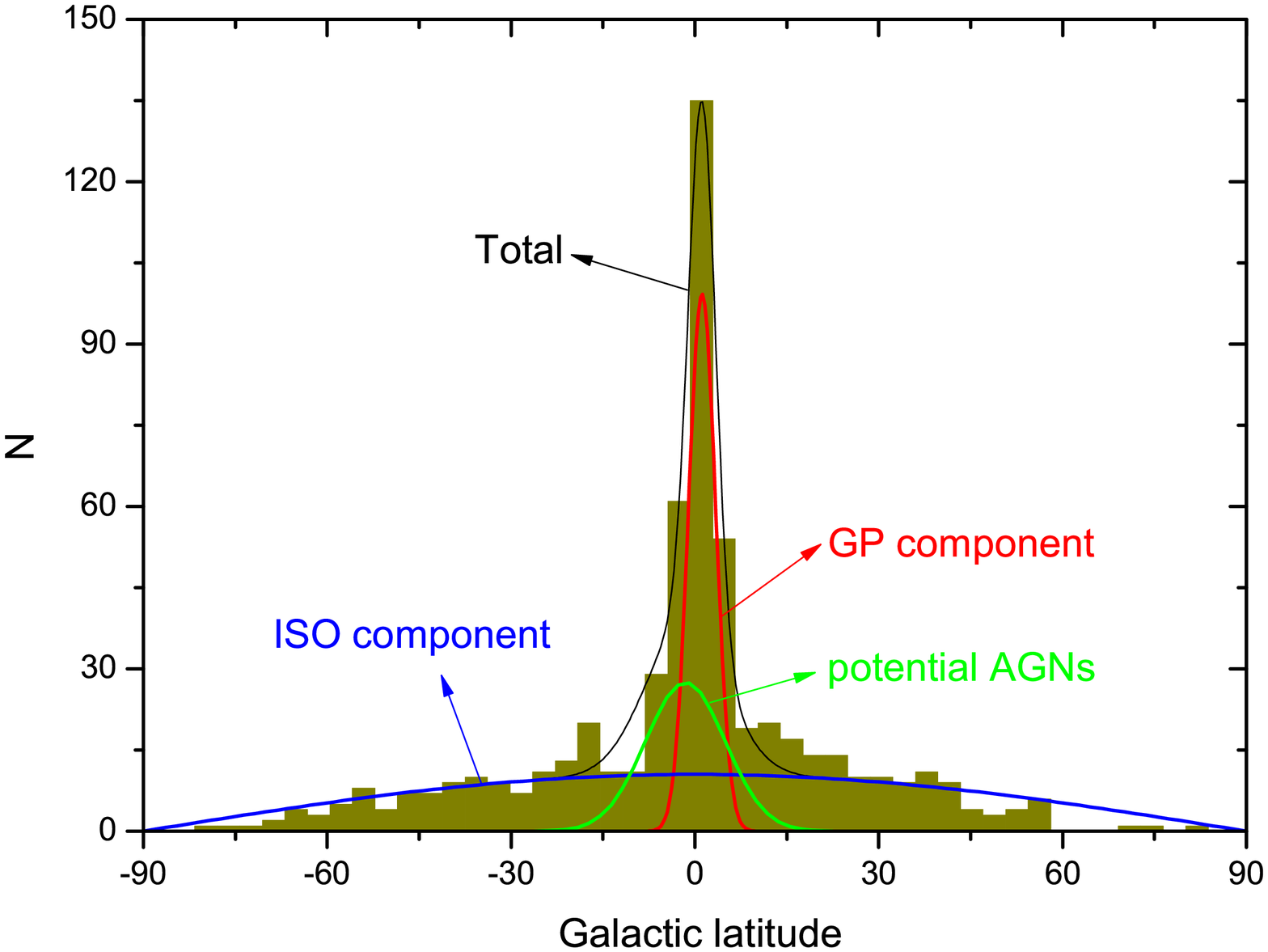}}
\caption{\textit{Left}: the spatial distribution of the 575
unassociated sources in the Galactic coordinates. \textit{Right}:
the latitude distribution of all unassociated sources and an
empirical fitting to the distribution with three components as
labeled.}
\end{figure}
\section{Sky distribution of the unassociated Fermi LAT sources}
The spatial distribution of the 575 unassociated Fermi LAT sources
is presented in the left panel of Figure 1, which exhibits an
obvious concentration of the sources in the Galactic plane (GP). In
more detail, we show the latitude distribution of these sources in
the right panel of Figure 1, where a spike in the source number
appears in the central $\sim10^{\circ}$ of the Galaxy. Such a spike
naturally indicates a Galactic origin for a remarkable fraction of
the unassociated sources. However, it should be noted that the spike
could also be partly contributed by some extragalactic active
galactic nuclei (AGNs) because of the anisotropic distribution of
the identified/associated AGNs. As shown in Figure 2, a big dip
appears in the AGN number distribution, which is caused by the
limited/no coverage of the AGN catalogs in the GP (Ackermann et al.
2012). Therefore, by considering the actual isotropic distribution
of AGNs, an extra component consisting of potential AGNs is expected
to exist in the unassociated sources. According to the fitting to
the latitude distribution of all identified/associated AGNs in
Figure 2, the number density of the potential AGNs per square degree
as a function of Galactic latitude can be empirically descried by
the following Gaussian function
\begin{eqnarray}
n_{\rm pAGN}(b)=n_{\rm pAGN,\max} e^{-(b-\mu_b)^2/2\sigma_b^2}
\end{eqnarray}
with $n_{\rm pAGN,\max}=0.02~{\rm deg^{-2}}$, $\mu_b=-1.6^{\circ}$,
and $\sigma_b=6.5^{\circ}$. Besides these potential AGNs, the
unassociated sources could still contain an
isotropically-distributed (ISO) component and a GP component. For a
simple empirical description, we express the number densities of the
ISO and GP components by a constant and a Gaussian function,
respectively. Then, as shown in the right panel of Figure 1, the
numbers of the unassociated sources within each $4^{\circ}$ bin as a
function of the latitude can be fitted by the following function
\begin{eqnarray}
N_{\rm bin}(b)=\left[n_{\rm GP,\max}
e^{-(b-\mu_{b})^2/2\sigma_{b}^2}+n_{\rm ISO}+n_{\rm
pAGN}(b)\right](2\pi \cos b\Delta b)
\end{eqnarray}
with $n_{\rm GP,\max}=0.07~{\rm deg^{-2}}$, $\mu_b=1.2^\circ$,
$\sigma_b=2.2^\circ$, and $n_{\rm ISO}=0.007~{\rm deg^{-2}}$, where
the factor $2\pi \cos b\Delta b$ represents the sky area of each
latitude bin.
\begin{figure}
\centering \resizebox{0.4\hsize}{!}{\includegraphics{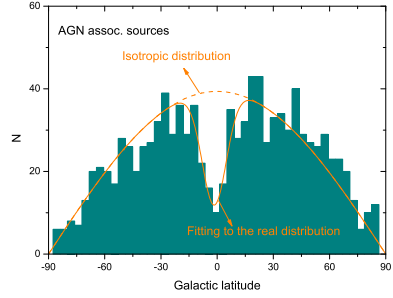}}
\caption{The latitude distribution of the sources associated with
AGNs, which is empirically fitted by the solid line. The dashed line
exhibits an isotropic distribution for a comparison. }
\end{figure}

Equation (2) indicates that the unassociated Fermi LAT sources can
be statistically separated into three components including the GP,
ISO, and potential AGN components, the fractions of which are about
$f_{\rm GP}=25\%$, $f_{\rm ISO}=55\%$, and $f_{\rm pAGN}=20\%$,
respectively. The fractions are obtained by integrating the
densities over the whole sky. Roughly speaking, one quarter of the
unassociated sources have Galactic origins, while the other three
quarters are possible extragalactic sources. Of course, such a
conclusion can somewhat be modified by some pulsars and some
possible exotic sources [e.g., due to dark matter annihilation
(Zechlin 2012)], which could locate in the Galactic halo and
contribute to the ISO component.


\begin{figure}
\resizebox{\hsize}{!}{\includegraphics{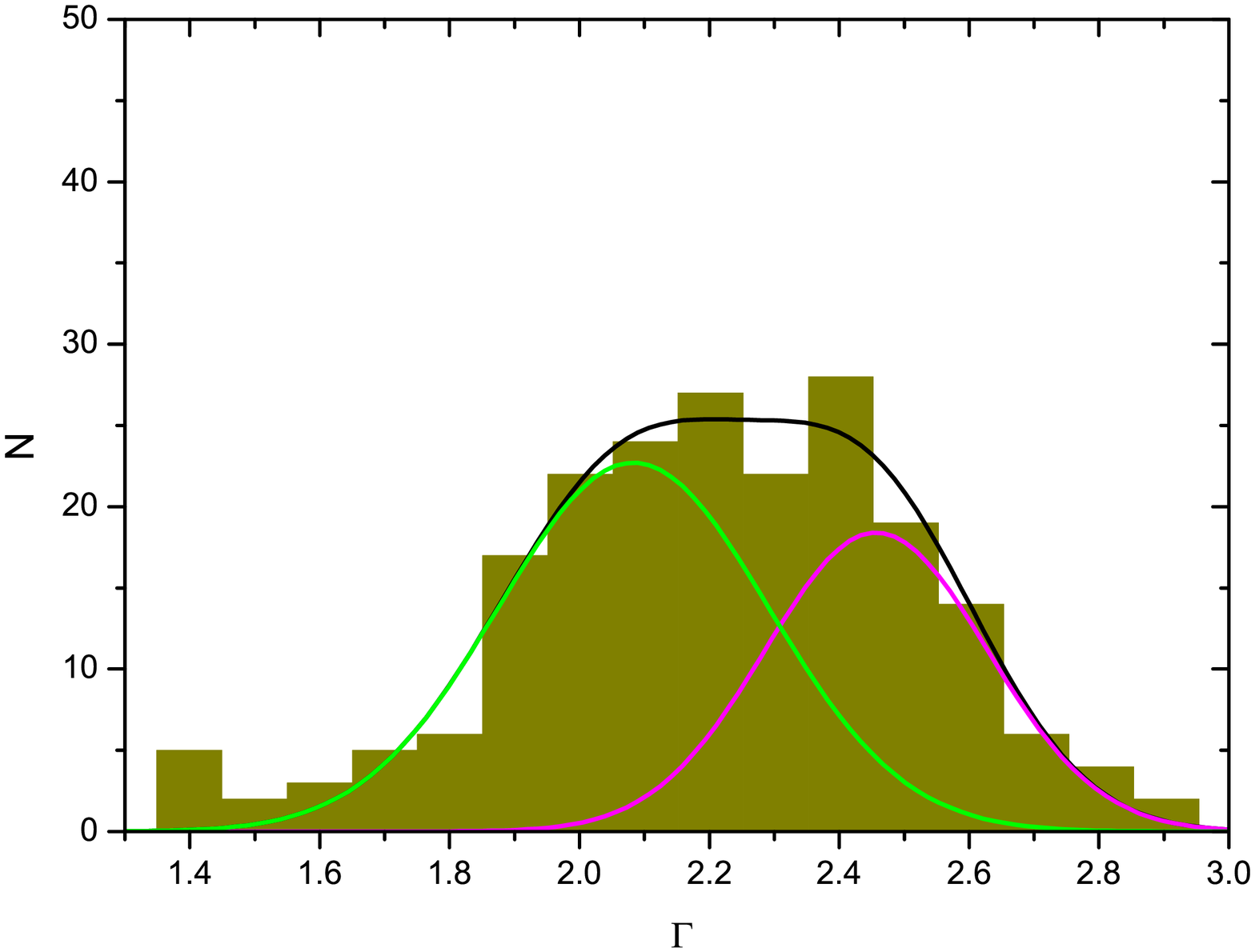}\includegraphics{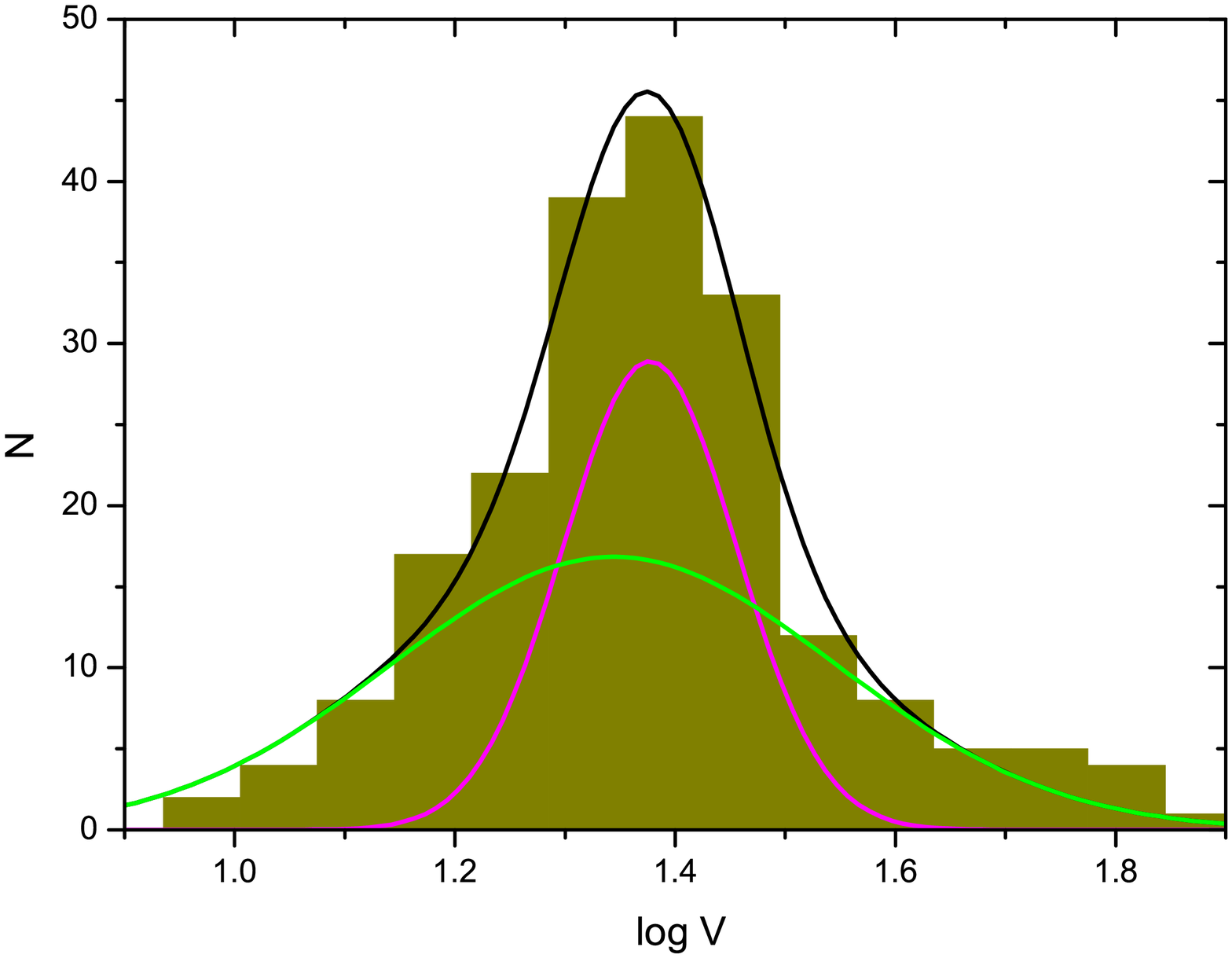}}
\caption{The distributions of the spectral and variability indices
of the 205 HL/ISO unassociated sources, both of which are fitted by
the sum of two Gaussians.}
\end{figure}
\begin{figure}
\resizebox{\hsize}{!}{\includegraphics{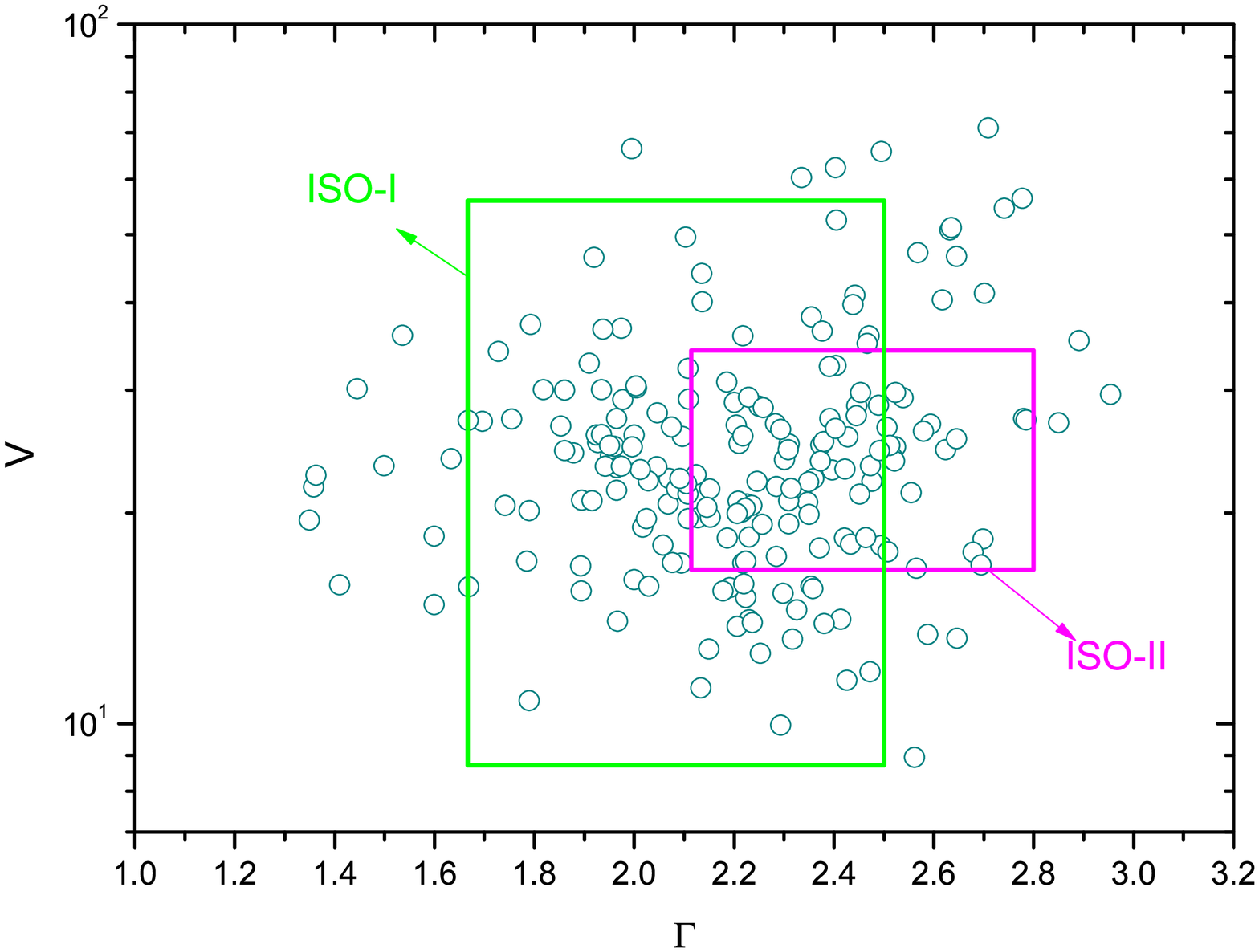}\includegraphics{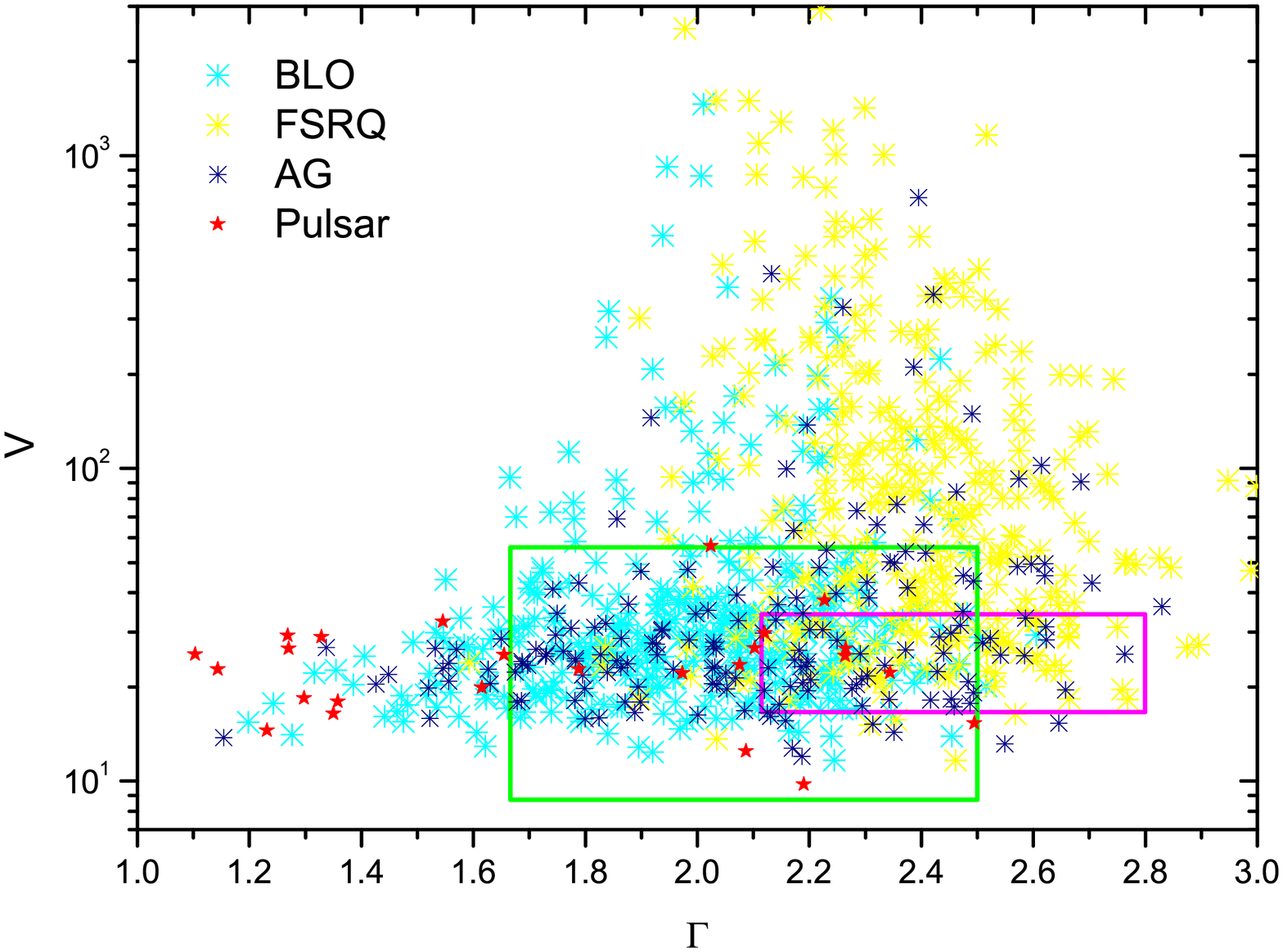}}
\caption{\textit{Left}: the $\Gamma-V$ distribution of the 205
HL/ISO unassociated sources, where the rectangles represent the
2-sigma regions of the ISO-I and ISO-II components. \textit{Right}:
A comparison between the 2-sigma regions of the ISO-I and ISO-II
components with the identified/associated sources at
$|b|\geq18^{\circ}$ including blazars, active galaxies, and
pulsars.}
\end{figure}
\section{Possible natures of the unassociated Fermi LAT sources}
As provided by the Fermi LAT collaboration (Abdo et al. 2010; Nolan
et al. 2012), the gamma-ray properties of the unassociated sources
can be characterized by four basic parameters as flux $S$, time
variability index $V$, spectral index $\Gamma$, and curvature index
$C$ which quantifies the departure of the spectrum from a single
power-law shape. In this paper, we make a basic assumption that a
quantity which reflects an intrinsic physical property of a source
type should exhibit a normal or lognormal distribution. However, due
to the distance-dependence of $S$ and the possible flux-dependence
of $C$ (Ackermann et al. 2012), these two parameters are not
considered to be good statistical quantities. Therefore, we would
only pay attention to the spectral and variability indices, $\Gamma$
and $V$, in our work.
\subsection{The ISO component}
\begin{table}
\centering
  \caption{Fitting parameters for distributions of $\Gamma$ and $V$}
 \begin{tabular}{cc|ccccccc}
      \hline\hline\noalign{\smallskip}
      & &\multicolumn{2}{c}{$\Gamma$}  &\multicolumn{2}{c}{$\log V$} \\  
\hline\noalign{\smallskip}
      &     &$\mu$ &     $\sigma$      &$\mu$      & $\sigma$    \\
\hline\noalign{\smallskip}
        &ISO-I    & $2.075$     & $0.204$   & $1.345$     & $0.187$    \\     
        &ISO-II    & $2.451$     & $0.174$   & $1.380$     & $0.073$    \\    
        &GP       & $2.440$     & $0.207$   & $1.364$     & $0.132$    \\     
\hline\hline
\end{tabular}
\end{table}

In order to reveal the possible nature of the ISO component, we
select the unassociated sources at latitudes higher than
$|b|=18^\circ$, the number of which is $N_{\rm HL}=205$. According
to the fitting given in the right panel of Figure 1, it can be
certain that all of these 205 high-latitude (HL) sources belong to
the ISO component. In Figure 3, we display the number distributions
of the spectral and variability indices of the 205 HL/ISO
unassociated sources by histograms. What these distributions tell
us? Following our basic assumption, the distributions of $\Gamma$
and $V$ are fitted simultaneously and tentatively by one, two, or
three Gaussians. By comparison, the best fit is obtained with two
Gaussians, as shown in Figure 3. The fitting function reads
\begin{eqnarray}
N_{\rm bin}={\mathcal{R}A\over\sigma_{\rm I}\sqrt{2\pi}}
e^{-(b-\mu_{\rm I})^2/2\sigma_{\rm I}^2}+{A\over\sigma_{\rm
II}\sqrt{2\pi}} e^{-(b-\mu_{\rm II})^2/2\sigma_{\rm II}^2},
\end{eqnarray}
and the fitting parameters are listed in Table 2 for $\mathcal
{R}=1.5$. Here we would like to emphasize that the ratio
$\mathcal{R}$ between the numbers of the two Gaussian components
should be the same in both the $\Gamma$ and $V$ fittings. According
to the above fittings, we can further separate the ISO component
into two subcomponents, denoted by ISO-I and ISO-II. Relative to the
total unassociated sources, the percentages of the ISO-I and ISO-II
components can be estimated to $f_{\rm ISO-I}=f_{\rm
ISO}{\mathcal{R}\over 1+\mathcal{R}}=33\%$ and $f_{\rm
ISO-II}=f_{\rm ISO}{1\over 1+\mathcal{R}}=22\%$, respectively.

In the left panel of Figure 4, we scatter the 205 HL/ISO
unassociated sources in the $\Gamma-V$ plane and show the 2-sigma
regions of the ISO-I and ISO-II components by the rectangles
according to Equation (3). As seen, most of the data is covered by
the two rectangles, except for the data on the most left and the top
right corner. In the right panel of Figure 4, nearly all of the
identified/associated sources at $|b|\geq 18^{\circ}$, including BL
Lacerate objects (BLOs), flat spectrum radio quasars (FSRQs), active
galaxies (AGs)\footnote{In a statistical view, we treat all of the
active galaxies with uncertain type, non-blazar active galaxies,
radio galaxies, Seyfert galaxies, starburst galaxies, and normal
galaxies with strong gamma-ray emission as active galaxies.}, and
pulsars, are displayed in comparison with the obtained 2-sigma
regions of the HL/ISO unassociated sources, which shows that: (i)
All of the sources with a large variability have been associated to
AGNs, and the V-range of the unassociated sources overlaps basically
the one of pulsars. (ii) The $\Gamma$-values of the HL/ISO
unassociated sources are in well agrement with the ones of most
AGNs, but somewhat higher than pulsars. At the same time, the offset
between the $\Gamma$-ranges of the BLOs and FSRQs is well reproduced
by the ISO-I and ISO-II components. Therefore, we suggest that the
nature of the ISO-I and ISO-II unassociated sources could be BLOs
and FSRQs, respectively. Finally, (iii) according to the
distribution of pulsars, in principle, it can not be ruled out that
some sources classified to the ISO-I component are actually pulsars.
However, the scarcity of unassociated sources with a hard spectrum,
which should be usual in pulsars, could indicate a low possibility
of pulsars for the unassociated sources. By ignoring the pulsar
possibility, the numbers of the unassociated sources as candidates
of BLOs and FSRQs can be estimated to
\begin{eqnarray}
N_{\rm BLO~cand}&=&N_{\rm unassoc}(f_{\rm ISO-I}+f_{\rm pAGN}f_{\rm BLO})=235,\\
N_{\rm FSRQ~cand}&=&N_{\rm unassoc}(f_{\rm ISO-II}+f_{\rm
pAGN}f_{\rm FSRQ})=165,
\end{eqnarray}
where $N_{\rm unassoc}=575$, and the fractions $f_{\rm BLO}=39\%$
and $f_{\rm FSRQ}=34\%$ are obtained by counting the numbers of the
identified/associated AGNs. Additionally, the remaining sources in
the potential AGNs could be AG candidates, the number of which is
about
\begin{eqnarray}
N_{\rm AG~cand}&=&N_{\rm unassoc}f_{\rm pAGN}(1-f_{\rm BLO}-f_{\rm
FSRQ})=31.
\end{eqnarray}

\subsection{The GP component}
\begin{figure}
\resizebox{\hsize}{!}{\includegraphics{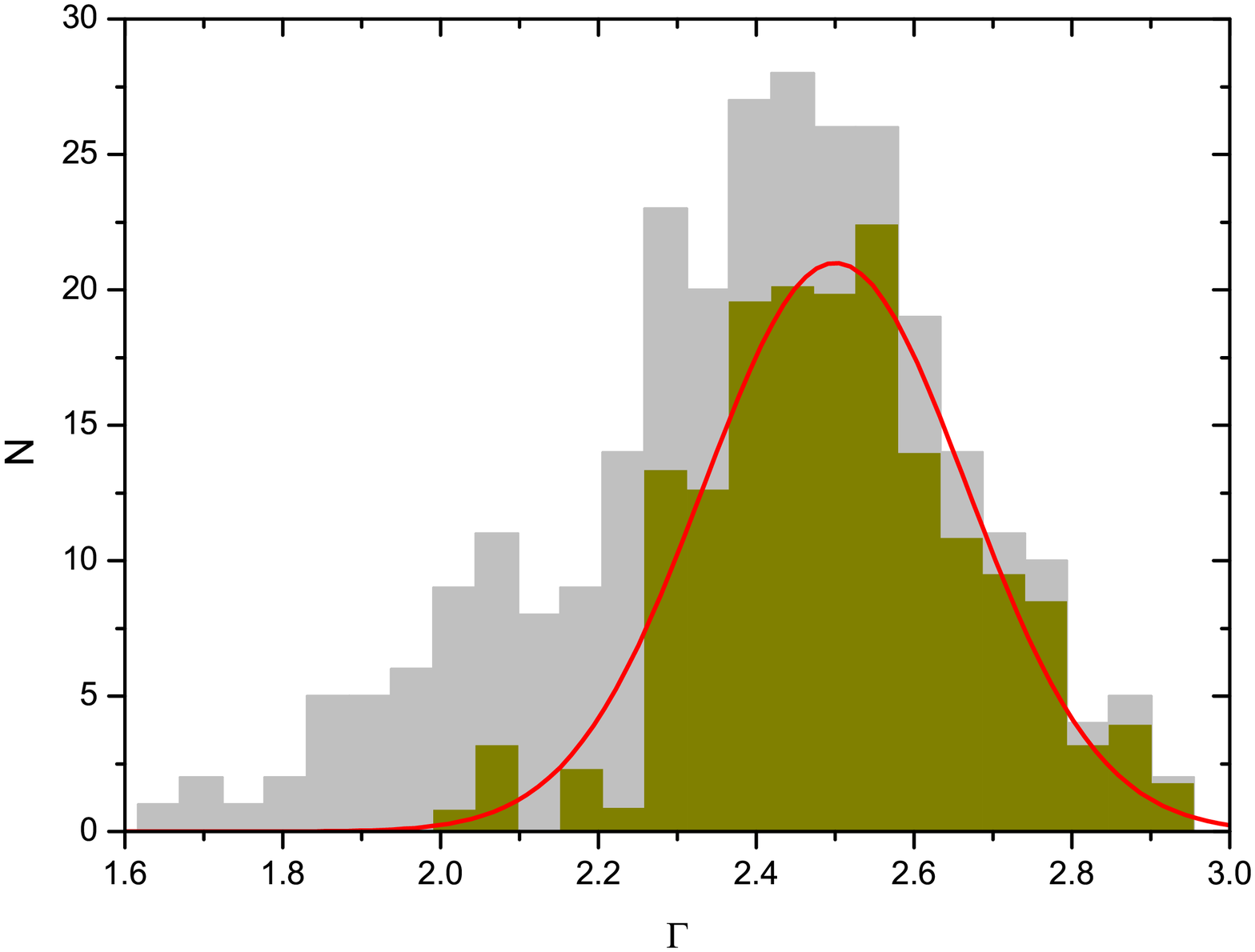}\includegraphics{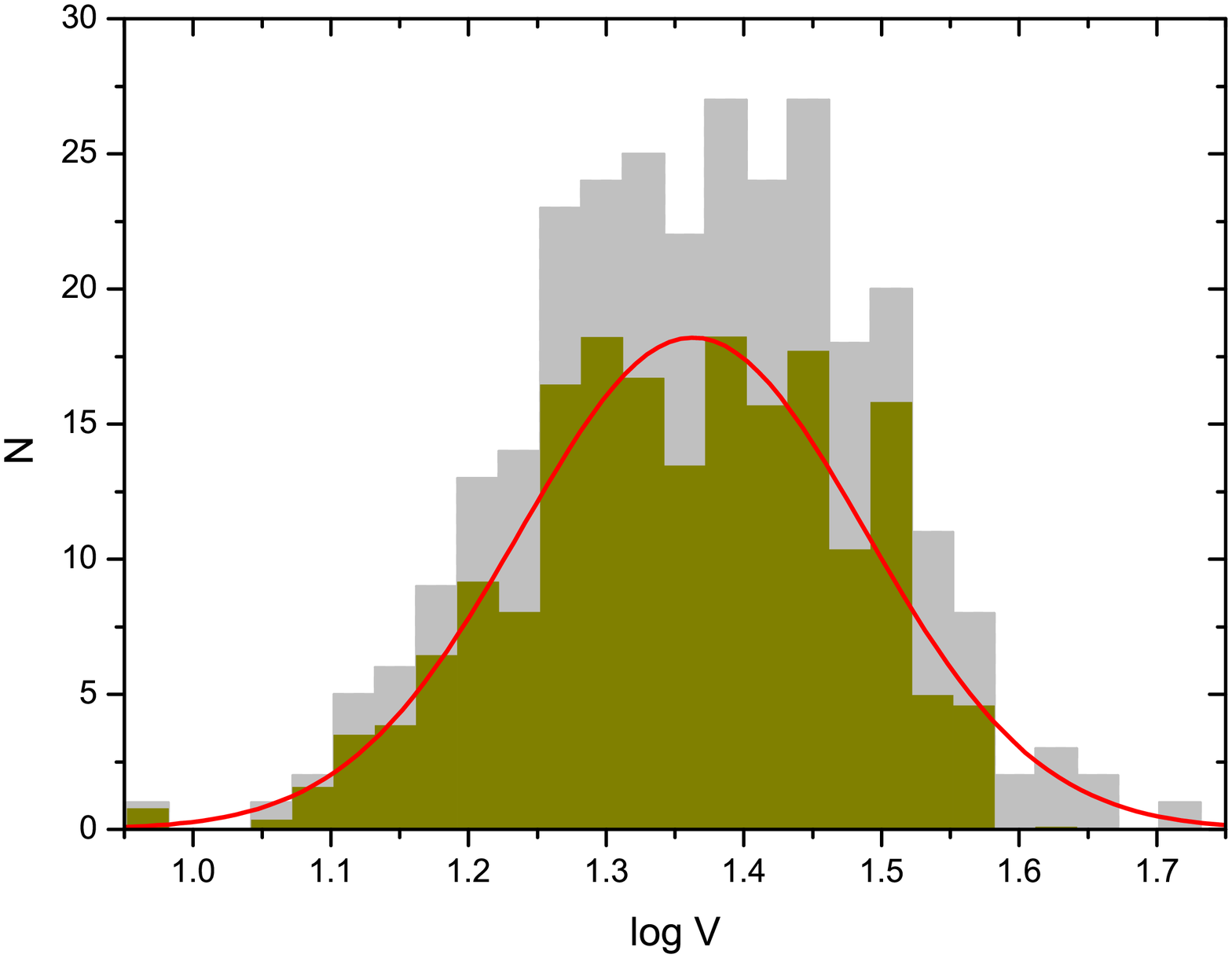}}
\caption{The distributions of the spectral and variability indices
of the 288 LL unassociated sources (grey histograms) and the
corrected distributions of only the GP component (dark yellow
histograms). The solid lines provide a single Gaussian fitting to
the GP sources.}
\end{figure}
\begin{figure}
\resizebox{\hsize}{!}{\includegraphics{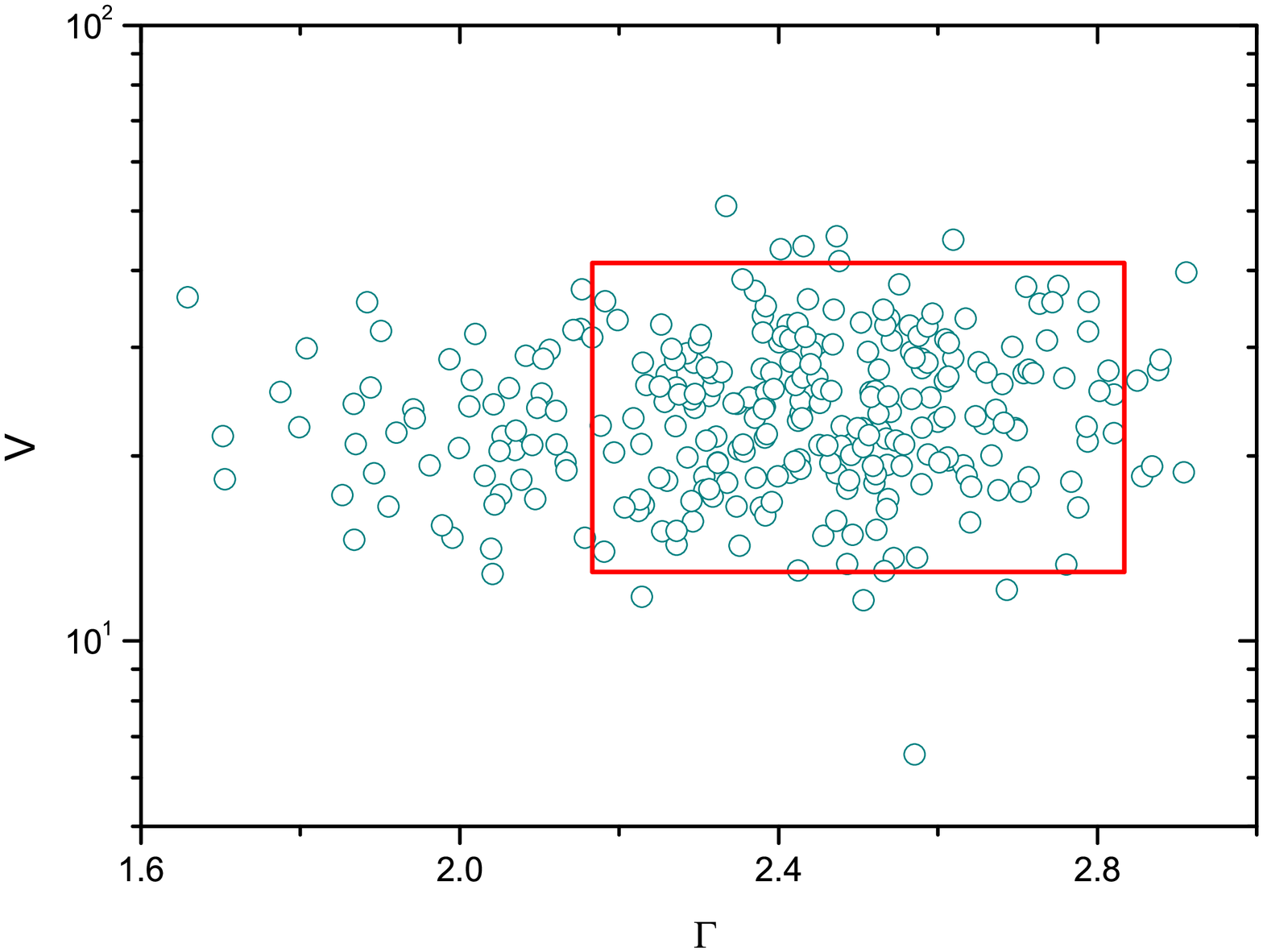}\includegraphics{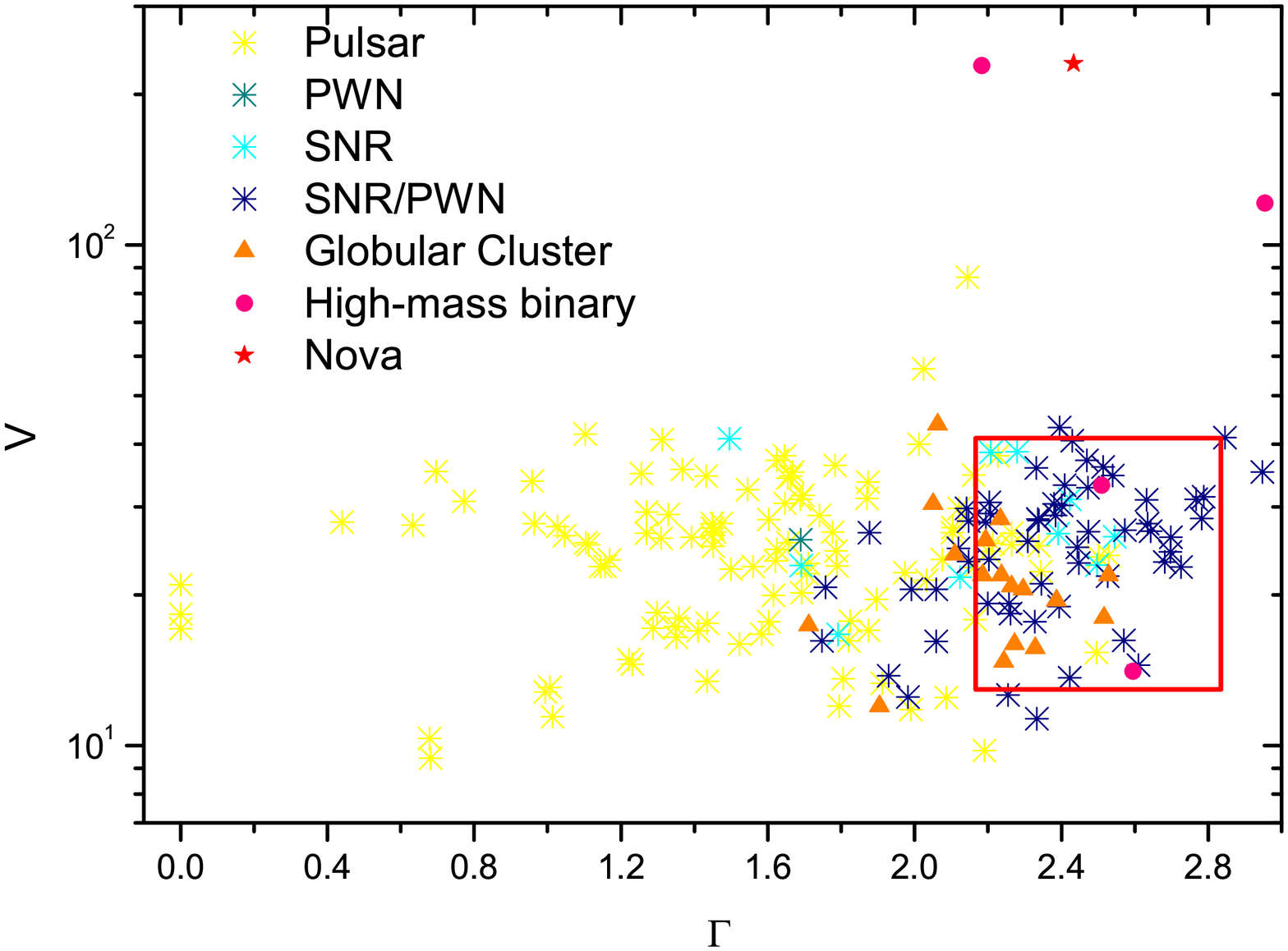}}
\caption{\textit{Left}: the $\Gamma-V$ distribution of the 288 LL
unassociated sources, where the rectangle represents the 2-sigma
regions of the GP component. \textit{Right}: A comparison between
the 2-sigma region of the GP component with the
identified/associated sources of Galactic origin including pulsars,
supernova remnants, and pulsar wind nebulae.}
\end{figure}
As a similar treatment, we select the unassociated sources at
latitudes lower than $|b|=8^{\circ}$, the number of which is $N_{\rm
LL}=288$. According to the fitting in the right panel of Figure 1,
we know that the low-latitude (LL) sources could be dominated by the
sources of Galactic origin. In Figure 5, we firstly display the
distributions of $\Gamma$ and $V$ of the 288 LL unassociated sources
by the grey histograms. However, these distributions can not
directly reflect the intrinsic properties of the GP component,
because a remarkable fraction of the LL sources are actually
extragalactic sources. According to Equation (2), we can get the
percentages of the different components in the latitude range of
$|b|\leq8^{\circ}$ as $f_{\rm LL,GP}=52\%$, $f_{\rm LL,ISO}=16\%$,
and $f_{\rm LL,pAGN}=32\%$. Therefore, the intrinsic distributions
of $\Gamma$ and $V$ of the GP sources can be obtained by the
following method
\begin{eqnarray}
N_{\rm GP,bin}=N_{\rm LL,bin}-f_{\rm ISO}N_{\rm ISO,bin}-f_{\rm
pAGN}N_{\rm pAGN,bin},
\end{eqnarray}
where $N_{\rm ISO,bin}$ and $N_{\rm pAGN,bin}$ are the numbers of
the ISO unassociated sources and potential AGNs in each $\Gamma$ or
$\log V$ bin. As a result, the corrected distributions of $\Gamma$
and $V$ of only the GP component are presented by the dark yellow
histograms in Figure 5, where two single-Gaussian fittings are
provided. The fitting parameters are also listed in Table 2. In the
left panel of Figure 6, we scatter the 288 LL unassociated sources
in the $\Gamma-V$ plane, where the 2-sigma region of the GP
component is presented by the rectangle. As seen, a remarkable
number of the data points locate beyond the rectangle, in
particular, at the low-$\Gamma$ range, which belongs to the ISO
sources and potential AGNs. In the right panel of Figure 6, we
compare the 2-sigma region of the GP component with the
identified/assocated GP sources including pulsars, SNRs, and PWNe.
The comparison suggests that most of the unassociated GP sources are
probably candidates of SNRs/PWNe, rather than usually expected
pulsars. The number of these GP sources is
\begin{eqnarray}
N_{\rm SN/PWN~cand}=N_{\rm unassoc}f_{\rm GP}=144.
\end{eqnarray}
Additionally, the GP component may also contain some globular
clusters, because most identified/associated globular clusters
locate in the bottom-left corner of the 2-sigma region although
their number is not very large.

\subsection{Identification efficiency}
\begin{figure}
\centering \resizebox{0.5\hsize}{!}{\includegraphics{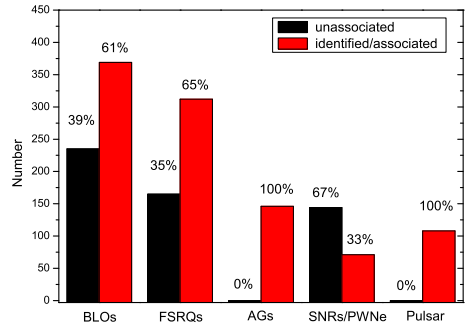}}
\caption{The percentages of the unassociated and
identified/associated sources of the types of BLOs, FSRQs, AGs,
SNRs/PWNe, and pulsars, where the cosmological sources are only
considered for $|b|\geq 18^{\circ}$ in order to avoid the
complication due to the potential AGNs.}
\end{figure}
In Figure 7 we present the identification efficiencies of BLOs,
FSRQs, AGs, SNRs/PWNe, and pulsars by comparing the numbers of
identified/associated sources with unassociated ones, where the
potential AGNs are not considered in order to avoid the complication
in the GP.

Firstly, the identification efficiencies of BLOs and FSRQs are
comparable to each other, whereas the efficiency of SNRs/PWNs is
much lower than the former two types. Such a situation could be
understood as follows. On one hand, most BLOs and FSRQs can easily
be distinct from other types of sources by their characteristic
rapid variability (large $V$). Moreover, the spectral energy
distributions of BLO and FSRQ emission usually peak at two energy
bands, i.e., the IR to X-Ray band and the MeV to TeV band. The
low-energy peak thus makes it relatively easy to find an IR (e.g.,
D'Abrusco et al. 2013; Massaro et al. 2013) or X-ray (e.g.,
ROMA-BZCAT sources; Massaro et al. 2009) counterpart. On the other
hand, for SNRs, two types of gamma-ray emission scenarios have been
widely investigated in literature, i.e, hadronic and leptonic
scenarios (Blandford and Ostriker 1978; Becker et al. 2011; Schuppan
et al. 2012). In the former case, gamma-ray emission is produced by
neutral pion decay, whereas in the latter case is by inverse-Compton
scattering of relativistic electrons on some seed photons. It could
be expected that, in the hadronic scenario, the low-energy emission
is probable negligible unless there are some other emission regions.
This makes it difficult to find low-energy counterparts for the
gamma-ray SNRs. In other words, the low identification efficiency of
SNRs may indicate that their gamma-ray emission is produced by
hadrons. Very recently, Mandelartz (2013) indeed claimed that
hadronic emission can be found in most Galactic SNRs. Future
searches of TeV photons (e.g., by H.E.S.S.) could be helpful to
distinguish these two scenarios, because in the leptonic scenario
the GeV and TeV photons have different origins.

Secondly, our statistical classification indicates that candidates
of pulsars and AGs in unassociated sources must be very limited. In
other words, the identification efficiencies of pulsars and AGs in
the Fermi LAT catalog are close to unity, except for the GP region
for AGs. Such high efficiencies could be due to (i) the plentiful
multi-wavelength observations of AGs which provide them sufficiently
enough low-energy counterparts and (ii) the apparent gamma-ray
pulsations of pulsars which significantly weakens the requirement of
other band counterparts for their identifications.

\section{Conclusions}

According to their spatial distribution, 575 unassociated Fermi LAT
sources can be separated empirically into GP, ISO and potential AGN
components, the fractions of which are 25\%, 55\%, and 20\%,
respectively. By comparing the spectral and variability properties
of the different classes of unassociation sources with
identified/associated sources, we conclude that the nature of the GP
component is probably SNRs/PWNe, while the ISO component can further
be separated into ISO-I and ISO-II subcomponents which are likely to
be associated with BLOs and FSRQs, respectively. To be specific, the
575 unassociated sources could statistically consist of $\sim$144
SNRs/PWNe, $\sim$235 BLOs, $\sim$165 FSRQs, and $\sim$31 AGs, where
the constituents of the potential AGNs in the GP are considered to
be the same to the identified/associated AGNs. The identification
efficiencies of BLOs, FSRQs, and SNRs/PWNe can further be estimated
to 61\%, 65\%, and 33\%, respectively, except for the GP region
where the efficiencies of BLOs and FSRQs become very low. In
contrast, our result indicates that the identification efficiencies
of pulsars and AGs could be very high. Anyway, the above conclusions
are only viable statistically. The existence of some exotic sources
can not be ruled out, because actually we can not determine the
nature of the unassociated sources individually. Nevertheless, the
statistical classification may still be helpful to a future
identification or association of the sources.


\begin{acknowledgements}
We thank Prof. K. S. Cheng for useful discussion which motivates
this work, Prof. S. N. Zhang for his instructive comments. This work
is supported by the National Natural Science Foundation of China
(Grant No. 11103004) and the Foundation for the Authors of National
Excellent Doctoral Dissertations of China (Grant No. 201225).
\end{acknowledgements}

\label{lastpage}

\end{document}